\documentclass[twocolumn,showpacs,preprintnumbers,amsmath,amssymb]{revtex4}

\usepackage{graphicx}
\usepackage{dcolumn}% Align table columns on decimal point
\usepackage{bm}% bold math

\usepackage{graphicx}
\usepackage{epsfig}

\begin{document}

\title{Nonsingular parametric oscillators Darboux-related to the classical harmonic oscillator}

\author{H.C. Rosu}
\email{hcr@ipicyt.edu.mx}
\affiliation{IPICyT, Instituto Potosino de Investigacion Cientifica y Tecnologica,\\
Apdo. Postal 3-74 Tangamanga, 78231 San Luis Potos\'{\i}, S.L.P., Mexico}

\author{O. Cornejo-P\'erez}
\email{octavio@uaq.mx}

\affiliation{Facultad de Ingenier\'{\i}a, Universidad Aut\'onoma de Quer\'etaro, Centro Universitario Cerro de las Campanas,
76010 Santiago de Quer\'etaro, Mexico}

\author{P. Chen}
\email{pisinchen@phys.ntu.edu.tw}

\affiliation{Leung Center for Cosmology and Particle Astrophysics and
Department of Physics and Graduate Institute of Astrophysics
National Taiwan University, Taipei, 10617 Taiwan}


\begin{abstract}
\noindent Interesting nonsingular parametric oscillators which are Darboux related to the classical harmonic oscillator and have periodic dissipative/gain features are identified through a modified factorization method. The same method is applied to the upside-down (hyperbolic) `oscillator' for which the obtained Darboux partners show transient underdamped features.
\end{abstract}

\maketitle

It is a common fact to factorize the classical harmonic oscillator equation $u''+\omega_{0}^{2}u=0$ through $\left( \frac{d}{dt }+R_{u_1}\right)\left( \frac{d}{dt}-R_{u_1}\right)u=0$, where $R_{u_1}=u'_1/u_1$ is the log-derivative of one of the solutions. We will use the harmonic mode $u_1=\cos\omega_0t$ in
the following. On the other hand, the equation \cite{rk}
%...1
\begin{equation}\label{d-osc}
v^{^{\prime\prime}} -\omega _{0}^{2}(2\tan ^2 \omega _0 t+1)v=0 %\qquad \omega_{v}^{2}(t) =-\omega _{0}^{2}(1+2\tan ^2 \omega _0 t)~.
\end{equation}
%..................................
is the result of the reversed factorization $\left( \frac{d}{dt}-R_{u_1}\right) \left( \frac{d}{dt }+R_{u_1}\right)v=0$.
%where $R_{u_1}$ is the Riccati solution obtained from the harmonic mode $u_1=\cos\omega_0t$.
 As well known, equation (\ref{d-osc}) is the Darboux-transformed equation of the oscillator equation. Since the frequency parameter is time dependent,
%$\omega_{v}$ depends on time,
it appears that the $v$ oscillator is parametric with a periodic chirping. Unfortunately, the linear independent modes of this oscillator
%...............
\begin{equation}\label{linindsol}
v_{1} \sim \frac{\omega _0}{\cos \omega _0 t}~, \qquad  v_{2} \sim \frac{1}{\omega _0\cos \omega _0 t}\left[\frac{\omega _0 t}{2}+\frac{1}{4}\sin 2\omega _0t\right]
%\qquad \rightarrow \qquad a_{1v}(\eta) \sim v_1^{1/\tilde{\gamma}}
~.
\end{equation}
are periodically singular which casts doubts on their possible applications.

In this Letter, we show that more general factorizations as discussed in \cite{1,2}, which we call $\alpha\beta$ factorizations herein, lead to a new class of Darboux type partner oscillators of the harmonic oscillator, which while being parametric are not singular. Since these factorizations can be applied to any second-order linear differential equation, we present them next using the common variable $x$ and then return to the time variables when we will discuss the simplest possible applications related to the harmonic and hyperbolic oscillators.

%\medskip
%
%{\bf The $\alpha\beta$ Factorizations}

\medskip

By $\alpha\beta$ factorizations we mean factorizations of second order linear differential operators performed with the following type of first-order operators:
%..............
%\begin{subequations}
\begin{eqnarray}\label{bmenos}
 B^- &= \alpha^{-1}(x)\frac{d}{dx}\pm\beta(x), \nonumber \\
B^+ &= \alpha(x)\frac{d}{dx}+\beta(x)~.
\end{eqnarray}
%\end{subequations}
%Of course, one can also use the following pair
%%..............
%%\begin{subequations}
%\begin{eqnarray}\label{bmenos1}
% B^- &= \alpha^{-1}(x)\frac{d}{dx}-\beta(x), \nonumber\\
%B^+ &= \alpha(x)\frac{d}{dx}+\beta(x)~.
%\end{eqnarray}
%\end{subequations}
The minus sign in the factorization operator $B^-$ produces only some sign changes in the rest of the mathematical formulas which are not important for the final results and therefore we will consider only the plus sign in the following.

\medskip

Then, we have:
%\begin{equation}\label{2}
%B^+B^-=(\alpha D+\beta)(\alpha^{-1}D+\beta)=D^2-\frac{\alpha'}{\alpha}D+\alpha\beta D +\frac{\beta}{\alpha}D+\alpha\beta'+\beta^2
%\end{equation}
%so that
\begin{equation}\label{2}
B^+B^-=D^2+\left(-\frac{\alpha _x}{\alpha}+\alpha\beta  +\frac{\beta}{\alpha}\right)D+(\beta^2+\alpha\beta_x)~,
\end{equation}
where $D=\frac{d}{dx}$ and the subindex $x$ also means the derivative with respect to $x$.

\bigskip

The reversed factorization reads:
%\begin{equation}\label{3}
%B^-B^+=(\alpha^{-1}D+\beta)(\alpha D+\beta)=D^2+\frac{\alpha'}{\alpha}D+\alpha\beta D +\frac{\beta}{\alpha}D+\frac{\beta'}{\alpha}+\beta^2
%\end{equation}
%so that
%6......................
\begin{equation}\label{3}
B^-B^+=D^2+\left(\frac{\alpha_x}{\alpha}+\alpha\beta  +\frac{\beta}{\alpha}\right)D+\left(\beta^2+\frac{\beta_x}{\alpha}\right)~.
\end{equation}

\bigskip

Thus, let us start with an equation of the form
\begin{equation}\label{ieq}
D^2y+f(x)Dy+g(x)y=0~,
\end{equation}
which for definiteness we assume to have only regular singular points in its coefficients. One can think of it to be a factorized product of either the first or the second form. However, in previous works \cite{1,2} it has been found that the second form is really tractable, while the first factorization leads to more complicated nonlinear equations for the factoring coefficients. Thus, considering (\ref{ieq}) to be of the $B^-B^+$ type, one is led to the following system of coupled equations:
%............
%\begin{subequations}
\begin{eqnarray}\label{a}
 & \alpha_x+\alpha^2\beta+\beta = \alpha f,\\ \label{b}
 & \beta^2+\frac{\beta_x}{\alpha} = g.
\end{eqnarray}
%\end{subequations}
This system of equations can be decoupled and leads to the following Riccati equation
%..............................
\begin{equation}\label{bern0}
-z_x-f(x)z+z^2+g(x)=0
\end{equation}
in the dependent variable $z=\beta/\alpha$. Thus, with one solution $h(x)$ of this Riccati equation at hand, one can get $\beta=h(x)\alpha$ and then (\ref{a}) turns into the following cubic Bernoulli equation
\begin{equation}\label{bern1}
\alpha_x+h(x)\alpha^3+[h(x)-f(x)]\alpha = 0~.
\end{equation}
The solution of this equation is related to the solution of the first-order differential equation
\begin{equation}\label{bern2}
w_x-2(h-f)w-2h=0
\end{equation}
through $\alpha=w^{-1/2}$, which leads to
\begin{equation}\label{bern3}
\alpha_{f,h}(x)=\frac{e^{-\int^x (h-f)dx'}}{\left(\lambda +\int^x 2h e^{-2\int^{x'} (h-f)dx}dx'\right)^{1/2}}~,%~,%\,
 %\quad \rightarrow
%\beta_{f,h}(x)=h(x)\alpha_{f,h}(x)~. %\frac{h(x)}{\left(\lambda_Be^{2\int^x (h-f)dx'}-1\right)^{1/2}}~.
\end{equation}
where $\lambda$ is an integration constant. Then $\beta_{f,h}(x)=h(x)\alpha_{f,h}(x)$ and the reversed-factorized equation is
%12
\begin{equation}\label{reveq}
D^2y+\left(-\frac{2\alpha_x}{\alpha}+f(x)\right)Dy+\left(g(x)+\beta_x(\alpha-\alpha^{-1})\right)y=0~.
 \end{equation}
This equation is the Darboux-transformed partner of equation (\ref{ieq}) and it will be of special interest in the following since it will be used to introduce a one-parameter family of nonsingular parametric oscillators related to the trigonometric and hyperbolic oscillators.

%\bigskip
%
%{\bf The classical harmonic oscillator case}

\medskip

 We move now to the simple but common case $f(x)=0$, which means equations originally without the first derivative or from which the first derivative has been eliminated according to a well known procedure. This is the standard case in quantum mechanics and has been addressed in the particular case of the quantum harmonic oscillator in \cite{1,2}. There is a slight difference produced by the minus sign taken in front of $\alpha$ in the operator $B^+$ which is included to get the negative sign in front of the kinetic term in the one-dimensional Schroedinger equation and also a scaling factor of $1/\sqrt{2}$ of the operators to get the $1/2$ numerical factor in front of the second derivative which is present for physical reasons.
The overall effect is a change of sign of the linear term in the Bernoulli equation (\ref{bern1}):
%13
\begin{equation}\label{bern4}
\alpha_x+h(x)\alpha^3-h(x)\alpha = 0~,
\end{equation}
with the solution easily obtained from (\ref{bern3}).
The quantum harmonic oscillator case corresponds to $h(x)=x$ \cite{1} if the particular solution is used, but one can also employ the general Riccati solution \cite{2}.

%\bigskip
%
%{\bf The classical harmonic oscillator case}

%\medskip

We focus now on the harmonic oscillator equation in classical mechanics, for which in addition to $f(t)=0$ one has $g(t)=const$, and we change the independent variable from $x$ to $t$. We have two non-trivial cases:\\

{\bf 1}. $g(t)=\omega_{0}^{2}$. $\quad$ This choice leads to the normal classical harmonic oscillator.
%and consider it to be of the $B^-B^+$ type.
In this case, we have \cite{rk}
%14...........................
\begin{equation}\label{ex2-1}
h(t)=\omega_0\tan\omega_0t~,
\end{equation}
where the initial condition of the Riccati equation is assumed to be included in the phase of the tangent function.
Then
%16..........................
\begin{equation}\label{ex2-2a}
\alpha=\pm \frac{\cos\omega_0 t}{\sqrt{\lambda-\cos^2\omega_0t}}~, \quad
\beta=\pm \frac{\omega_0 \sin\omega_0t}{\sqrt{\lambda-\cos^2\omega_0t}}~.
\end{equation}
We use the plus signs in the following since the minus ones do not produce any change in the results.

\begin{figure}[x]%[!h]
  \centering
  \includegraphics[width= 7.08 cm, height=9.07 cm]{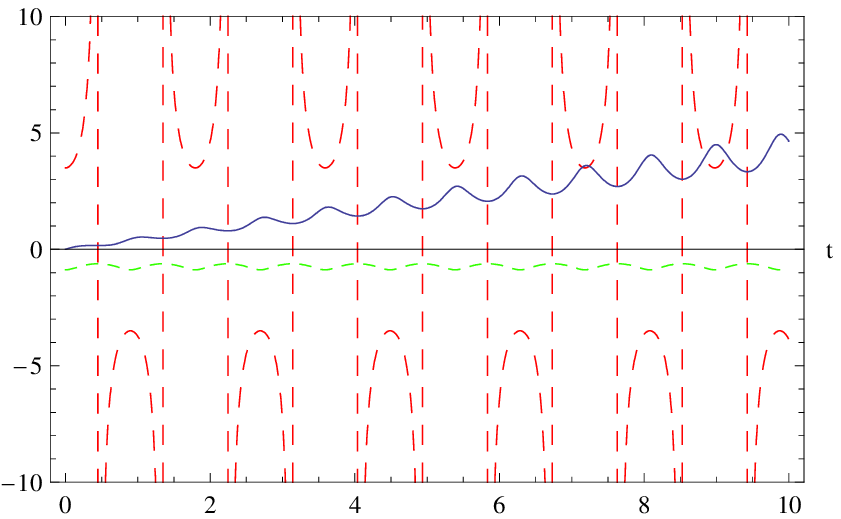}
  \caption{(Color online) Plot of the real part (blue line) and imaginary part (green line) of the solution (\ref{ex2-3b}) and of $v_1$ (red line) for $\lambda=2$, $C_1 = 2/7$, $C_2 =7/4$, and $\omega_0=3.5$.}
  \label{fig-ry1}
 \end{figure}

\begin{figure}[x]%[!h]
  \centering
  \includegraphics[width= 7.08 cm, height=9.07 cm]{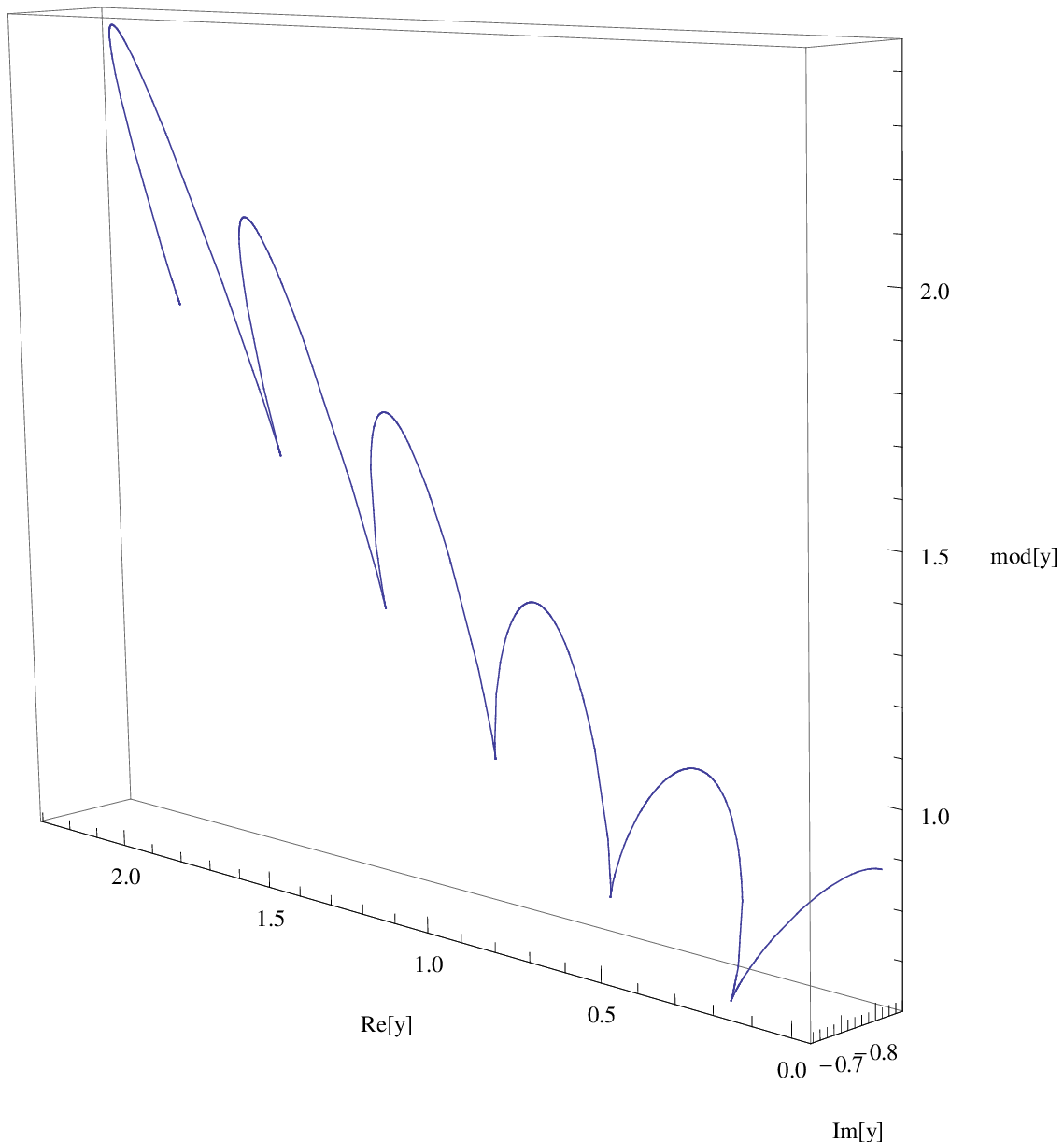}
  \caption{(Color online) Three-dimensional plot of the modulus of the solution (\ref{ex2-3b}) as a function of its real and imaginary parts for the same values of the parameters.}
  \label{fig-ry-m}
 \end{figure}

\begin{figure}[x]%[!h]
  \centering
  \includegraphics[width= 7.08 cm, height=9.07 cm]{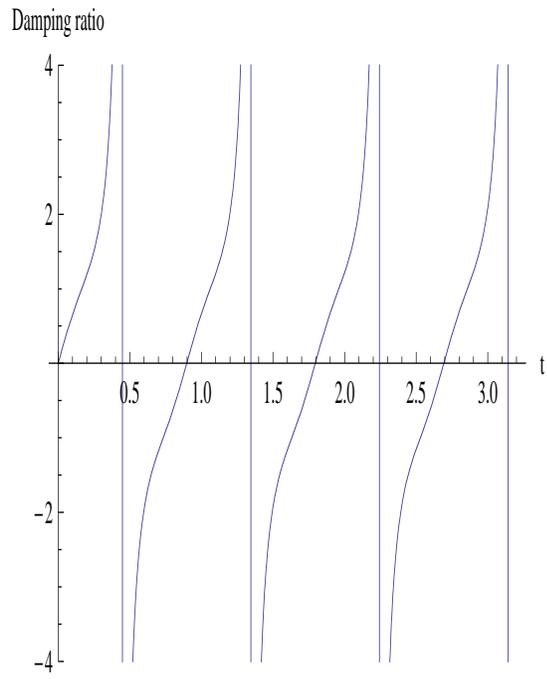}
  \caption{(Color online)  The damping ratio $\zeta_o(t)$ in (\ref{ex2-3abis}) for $\lambda=2$ and $\omega_0=3.5$.}
  \label{fig-dampt}
 \end{figure}

\begin{figure}[x]%[!h]
  \centering
  \includegraphics[width= 7.08 cm, height=9.07 cm]{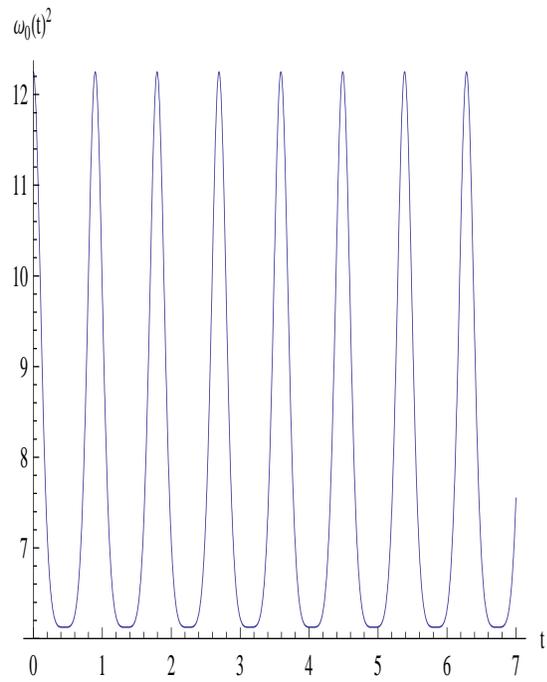}
  \caption{(Color online) Plot of $\omega_0^2(t)$ given by equation (\ref{ex2-3atris}) for $\lambda=2$ and $\omega_0=3.5$.}
  \label{fig-OmegaT}
 \end{figure}

The corresponding $B^+B^-y=0$ equation, i.e., $y'' -\left(\frac{2\alpha'}{\alpha}\right)y'+\left(\omega_{0}^{2}+\beta'(\alpha-\alpha^{-1})\right)y=0$, takes the form
%\begin{equation}\label{ex2-3}
%D^2y -\left(\frac{2\alpha'}{\alpha}\right)Dy+\left(\omega_{0}^{2}+\beta'(\alpha-\alpha^{-1})\right)y=0~.
%\end{equation}
%or:
%17...............................
\begin{equation}\label{ex2-3a}
y''+2\zeta _o(t)\omega_0y'+\omega_{0}^{2}(t)y=0~.
%\bigg[\frac{1}{\lambda-\cos^2\omega_0t}-\frac{\sin^2 2\omega_0t}{4(\lambda-\cos^2\omega_0t)^2}\bigg]y=0~.
\end{equation}
In the coefficient of the first derivative, we identify
%18...............................
\begin{equation}\label{ex2-3abis}
\zeta _o(t)=\frac{\lambda\tan\omega_0t}{\lambda-\cos^2\omega_0t}
\end{equation}
as the so-called damping ratio \cite{wiki-dr}, which is a function of time, differently from the common damped oscillators for which it is just a number in the interval $[0,1]$.
On the other hand,
%19............
\begin{equation}\label{ex2-3atris}
\omega_0^2(t)=\omega_{0}^{2}\bigg[\frac{1}{\lambda-\cos^2\omega_0t}
-\frac{\sin^2 2\omega_0t}{4(\lambda-\cos^2\omega_0t)^2}\bigg]
\end{equation}
is the time periodic frequency of this parametric oscillator.
The solution of (\ref{ex2-3a}) is:
%20.............................
\begin{equation}\label{ex2-3b}
y(t)=C_1\frac{\omega_0t+\frac{1}{2}\sin 2\omega_0t}{2\sqrt{\lambda-\cos^2\omega_{0}t}}-iC_2\frac{1}{2\sqrt{\lambda-\cos^2\omega_{0}t}}~.
\end{equation}
Notice that this solution does not have singularities if $\lambda\notin [0,1]$. Another interesting fact is that if one chooses $C_1=\frac{1}{\omega _0}$ and $C_2=\omega_0/2$, one can write this solution in the form
%21..............................
\begin{equation}\label{ex2-3c}
y(t)=-\frac{\cos\omega_0t}{\sqrt{\cos^2\omega_{0}t-\lambda}}(v_1(t)+iv_2(t))~,
%-i\frac{v_1(t)\cos\omega_0(t+t_0)}{\sqrt{\lambda-\cos^2\omega_{0}(t+t_0)}}
\end{equation}
which shows the connections with the $v$ oscillators. Plots of the real and imaginary parts of (\ref{ex2-3b}) for $\lambda=2$ are given in Fig.~\ref{fig-ry1} and of its modulus in Fig.~\ref{fig-ry-m}, while $\zeta_o(t)$ and $\omega_0^2(t)$ are displayed in Fig.~\ref{fig-dampt} and Fig.~\ref{fig-OmegaT}, respectively.
It is worth mentioning the fact that the friction coefficient in (\ref{ex2-3a}) is negative in a periodic manner. In general, the friction coefficient can be negative only as a result of some energy pumping mechanism. Such cases are known for example in plasma physics, especially for dusty plasmas \cite{dusty-p}, but one can think also of negative differential resistance circuits as well as mesoscopic and nanoscopic transport processes.\\

{\bf 2}. $g(t)=-k_{0}^{2}$. $\quad$ This case corresponds to the upside-down harmonic oscillator in the following sense.
The Riccati solution of (\ref{bern0}) is
%22..........................
\begin{equation}\label{ex2.2-1}
h(t)=-k_0\tanh k_0t~.
\end{equation}
For the standard factorization, one gets
%23........
\begin{equation}\label{ex2.2-1bis}
{\rm u}^{\prime\prime}-k _{0}^{2}(2\tanh^2k_0t-1){\rm u}=0
\end{equation}
 as initial equation having the linearly independent solutions
%24...............
\begin{equation}\label{linindsol2}
{\rm u}_{1} \sim \frac{k _0}{\cosh k_0 t}~, \qquad  {\rm u}_{2} \sim \frac{1}{k _0\cosh k _0 t}\left[\frac{k_0 t}{2}+\frac{1}{4}\sinh 2k _0t\right]~.
\end{equation}
The Darboux partner equation is
%25......
\begin{equation}\label{d-osc-h}
w^{^{\prime\prime}} -k _{0}^{2}w=0
\end{equation}
with the well-known linear independent zero-modes
%21...............
%\begin{equation}\label{linindsol-h}
$w_{1} \sim \cosh k _0 t$ and $w_{2} \sim\sinh k _0t$.
%\qquad \rightarrow \qquad a_{1v}(\eta) \sim v_1^{1/\tilde{\gamma}}
%~.\end{equation}

The $\alpha\beta$ factorization coefficients are now
%26.............................
\begin{equation}\label{ex2.2-2}
\alpha=\pm \frac{\cosh k_0t}{\sqrt{\lambda-\cosh^2k_0t}}~, \quad
\beta=\mp \frac{k_0 \sinh k_0t}{\sqrt{\lambda-\cosh^2 k_0t}}~.
\end{equation}
%The change of signs do not affect the results and we use the first pair.
The $B^+B^-y=0$ equation reads
%%27.............................
%\begin{eqnarray}\label{ex2.2-3}
\begin{equation}\label{ex2.2-3}
y''+2\zeta_h(t) k_0y'+k_{0}^{2}(t)y=0~,
%\left(\frac{1}{\cosh^2k_0t-\lambda}-\frac{\sinh 2k_0t}{4(\cosh^2k_0t-\lambda)^2}\right)y=0~.
\end{equation}
where the damping ratio is now
%28..........
\begin{equation}\label{dratio}
\zeta_h(t)=\frac{\lambda \tanh k_0t}{\cosh^2k_0t-\lambda}
\end{equation}
and the parametric frequency-like coefficient is
%29........
\begin{equation}\label{kt}
k_0^2(t)=k_{0}^{2}\left(\frac{1}{\cosh^2k_0t-\lambda}-\frac{\sinh 2k_0t}{4(\cosh^2k_0t-\lambda)^2}\right)~.
\end{equation}

\begin{figure}[x]%[!h]
  \centering
  \includegraphics[width= 7.08 cm, height=9.07 cm]{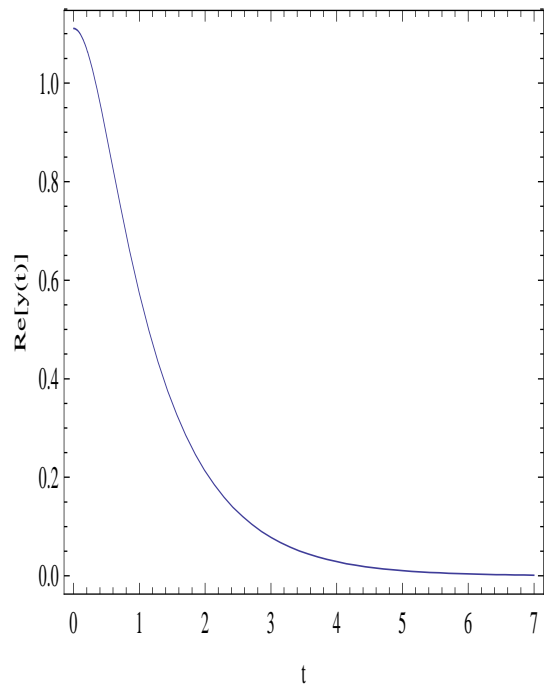}
  \caption{(Color online) The real part of the solution (\ref{ex2.2-4}) for the superposition constants $C_3=2$ and $C_4=-1$ and $k_0=1$ and $\lambda=0.5$.}
  \label{fig-ry2}
 \end{figure}

\begin{figure}[x]%[!h]
  \centering
  \includegraphics[width= 7.08 cm, height=9.07 cm]{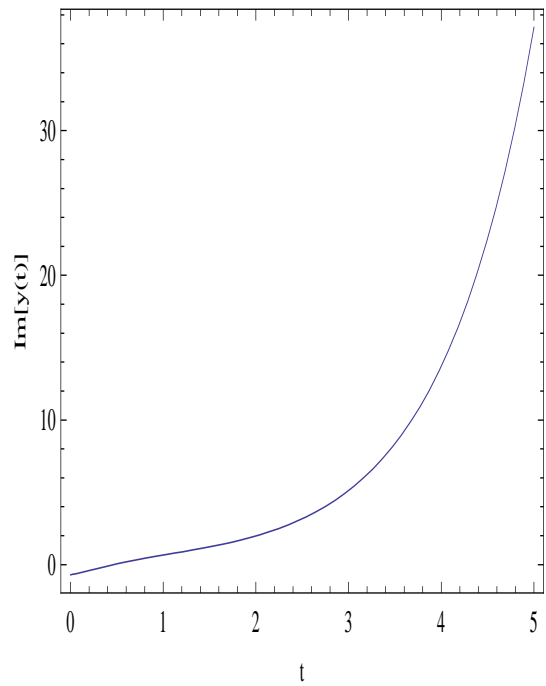}
  \caption{(Color online) The imaginary part of the solution (\ref{ex2.2-4}) for the same values of the parameters.}
  \label{fig-iy2}
 \end{figure}

\begin{figure}[x]%[!h]
  \centering
  \includegraphics[width= 7.08 cm, height=9.07 cm]{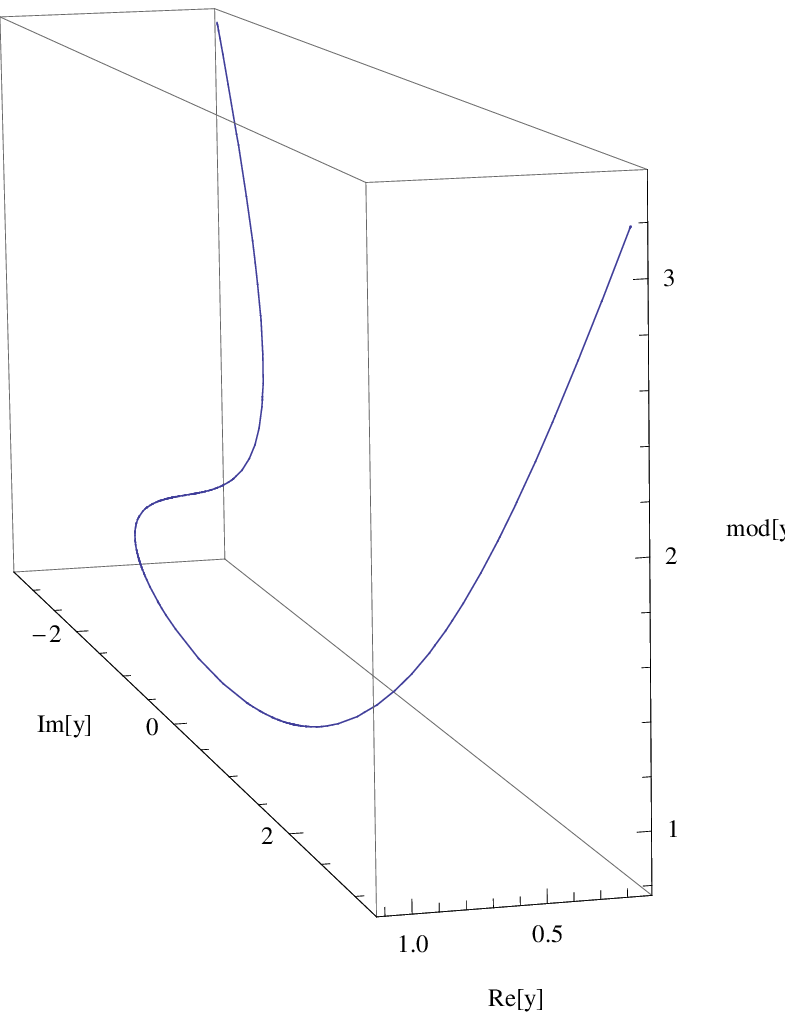}
  \caption{(Color online) Three-dimensional plot of the modulus of the solution (\ref{ex2.2-4}) as a function of its real and imaginary parts for the same values of the parameters.}
  \label{fig-y2mod}
 \end{figure}

\begin{figure}[x]%[!h]
  \centering
  \includegraphics[width= 7.08 cm, height=9.07 cm]{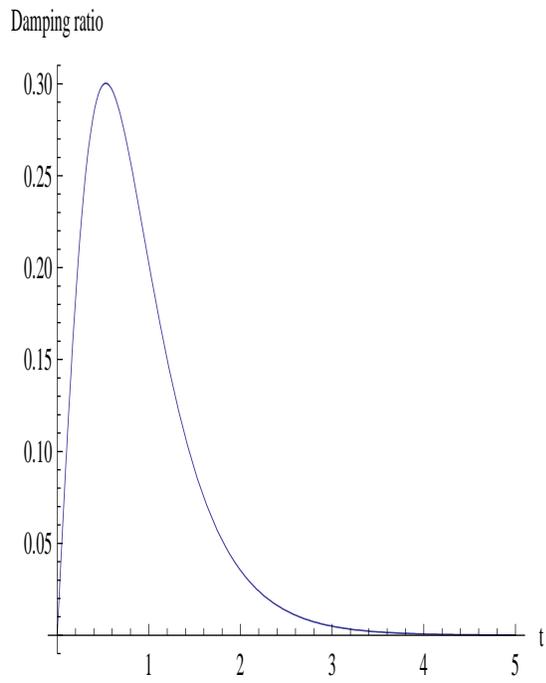}
  \caption{(Color online) The damping ratio $\zeta_h(t)$ in (\ref{dratio}) for $k_0=1$ and $\lambda=0.5$.}
  \label{fig-dampr}
 \end{figure}

\begin{figure}[x]%[!h]
  \centering
  \includegraphics[width= 7.08 cm, height=9.07 cm]{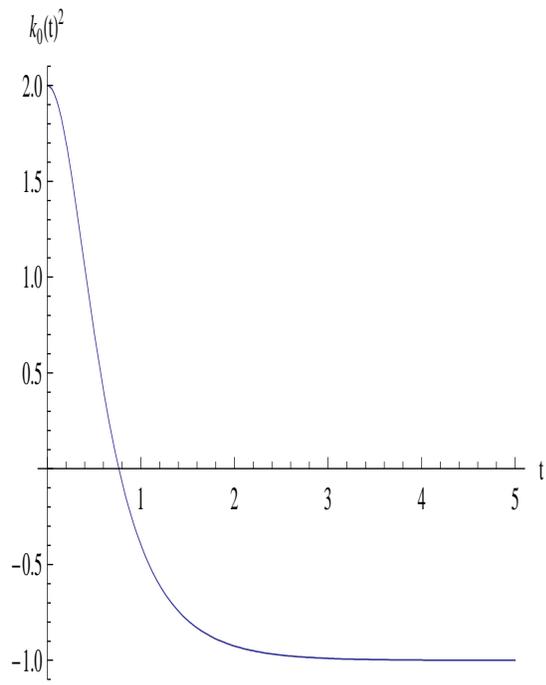}
  \caption{(Color online) Plot of $k_{0}^{2}(t)$ given by equation (\ref{kt}) for $\lambda=0.5$ and $k_0=1$.}
  \label{fig-OmegaH}
 \end{figure}

The general solution has the following explicit form
%30.............................
 \begin{equation}\label{ex2.2-4}
y(t)=\frac{[(C_3+iC_4\pi)+C_4(2k_0t+\sinh 2k_0t)]}{4\sqrt{\lambda-\cosh^2k_{0}t}}~.
%\left(C_3+C_4(2k_0t+\sinh 2k_0t)+iC_4\pi\right)~,
\end{equation}
If $\lambda <1$, no singularities occur in (\ref{ex2.2-2})-(\ref{ex2.2-4}).
Plots of the real and imaginary parts of the solution (\ref{ex2.2-4}) are shown in Figs.~(\ref{fig-ry2}) and (\ref{fig-iy2}), respectively, and of its modulus in Fig.~(\ref{fig-y2mod}). We also plot the damping ratio, $\zeta_h(t)$, in Fig.~(\ref{fig-dampr}) and the frequency-like coefficient $k_{0}^{2}(t)$ in  Fig.~(\ref{fig-OmegaH}), all of them for $\lambda=0.5$ and $k_0=1$. Examining these plots, it can be seen that after an initial period when both quantities are positive, the  damping ratio drops to zero, while $k_{0}^{2}(t)$ tends to $-k_{0}^{2}$. In other words, the parametric equation (\ref{ex2.2-3}) turns into the standard hyperbolic equation (\ref{d-osc-h}).

%...FIG 1

%... FIG 2

 %... FIG 3

 %... FIG 4

%... FIG 5

%... FIG 6

%... FIG 7

%... FIG 8

 %... FIG 9
 
 We finally notice that if we write the $B^+B^-$ equation as
\begin{equation}\label{FG1}
D^2y+F(x)Dy+G(x)y=0~,
\end{equation}
then
\begin{equation}\label{FG2}
F=f-2\frac{\alpha_x}{\alpha}~, \quad G=g+\beta_x\left(\frac{\alpha^2-1}{\alpha}\right)~.
\end{equation}
Thus, if $\alpha=\pm 1$ as in the standard factorization no new equation is obtained. On the other hand, it is easy to check that for $f=0$ the case $\alpha=\pm i$ leads to the undamped but singular parametric oscillators obtained by the standard factorization for $g=\pm$ const., which also correspond to taking $\lambda=0$. More generally, in the framework of the $\alpha\beta$ factorization, one can get undamped parametric equations only if $f=2\frac{\alpha_x}{\alpha}$, which leads to $F=0$ and $G=g+\left(\frac{f}{2}h+h_x\right)\left(e^{\int^x fdx}-1\right)$.

\bigskip

In summary, we have presented here the general method of $\alpha\beta$ factorizations and based on it we introduced an interesting family of nonsingular parametric oscillators which are Darboux related to the classical harmonic oscillators. The scheme has been also applied to the hyperbolic case.

\bigskip

Acknowledgment: HCR thanks CONACyT for a sabbatical fellowship. PC appreciates the supports by Taiwan National Science Council(NSC) under Project No. NSC-100-2119-M-002-525, No. NSC-100-2112-M-182-001-MY3, US Department of Energy under Contract No. DE-AC03-76SF00515, and the NTU Leung Center for Cosmology and Particle Astrophysics (LeCosPA).

%
%%
%%\begin{thebibliography}{123}
%%
%
%%%....................
%%\begin{eqnarray}\label{ficond}
%%-\left( \phi_1+\phi_2+\frac{d\phi_1}{du} \,u \right) &=& g(u)\\
%%\phi_1\,\phi_2 &=& \frac{F(u)}{u}~.
%%\end{eqnarray}
%%%....................

\end{document}